\documentclass[times, twoside]{StyleArXiv}
\usepackage{booktabs}
\usepackage{multicol,multirow}
\usepackage{orcidlink}
\usepackage{subcaption}
\usepackage{cleveref}
\crefname{figure}{Fig}{Figs}
\crefname{table}{Tab}{Tabs}
\crefname{equation}{eq.}{eqs.}

\leadauthor{Siddhad} 

\begin{document}

\title{Enhancing EEG Signal-Based Emotion Recognition with Synthetic Data: Diffusion Model Approach
\thanks{\copyright 2025 IEEE. Personal use of this material is permitted. Permission from IEEE must be obtained for all other uses, in any current or future media, including reprinting/republishing this material for advertising or promotional purposes, collecting new collected works for resale or redistribution to servers or lists, or reuse of any copyrighted component of this work in other works. [\href{https://doi.org/10.1109/TAI.2025.3641576}{10.1109/TAI.2025.3641576}]}}
\shorttitle{Enhancing EEG Signal-Based Emotion Recognition with Synthetic Data: Diffusion Model Approach}

\author[1,\Letter]{Gourav~Siddhad~\orcidlink{0000-0001-5883-3863}}
\author[2]{Masakazu~Iwamura~\orcidlink{0000-0003-2508-2869}}
\author[1]{Partha~Pratim~Roy~\orcidlink{0000-0002-5735-5254}}
\affil[1]{Department of Computer Science and Engineering, Indian Institute of Technology, Roorkee, Uttarakhand, 247667, India}
\affil[2]{Department of Core Informatics, Graduate School of Informatics, Osaka Metropolitan University, Sakai, Osaka, 599-8531, Japan}

\maketitle



\begin{abstract}
Emotions are crucial in human life, influencing perceptions, relationships, behaviour, and choices. Emotion recognition using Electroencephalography (EEG) in the Brain-Computer Interface (BCI) domain presents significant challenges, particularly the need for extensive datasets. This study aims to generate synthetic EEG samples similar to real samples but distinct by augmenting noise to a conditional denoising diffusion probabilistic model, thus addressing the prevalent issue of data scarcity in EEG research. The proposed method is tested on the DEAP and SADT datasets, showcasing up to 5.6\% improvement in classification accuracy when using synthetic data with DEAP and similar positive results with SADT. This is higher compared to the traditional Generative Adversarial Network (GAN) based and Denoising Diffusion Probabilistic Model (DDPM) based approaches. This study further evaluates the effectiveness of state-of-the-art classifiers on EEG data, employing both real and synthetic data with varying noise levels, and utilizes t-SNE and SHAP for detailed analysis and interpretability. The proposed diffusion-based approach for EEG data generation appears promising in refining the accuracy of emotion recognition systems and marks a notable contribution to EEG-based emotion recognition.
\end{abstract}


\begin{corrauthor}
g\_siddhad\at cs.iitr.ac.in
\end{corrauthor}


\section{Introduction}
\label{sec_intro}

Emotions are fundamental to human experience, influencing perceptions, relationships, behaviors, and decision-making. They significantly impact physical and mental health, mood, energy levels, and overall well-being. Understanding and responding effectively to emotions is crucial for both interpersonal interactions and personal success. Electroencephalography (EEG), a non-invasive and cost-effective neuroimaging technique, has emerged as a promising tool for capturing subtle changes in emotional states. It has been employed in Brain-Computer Interfaces (BCIs) for real-time responses to affective states~\cite{chakladar2021eeg}. EEG has shown potential applications in human-computer interaction, emotion recognition~\cite{houssein2022human}, and medical diagnosis~\cite{liu2021review}. However, EEG-based emotion classification remains challenging despite its high temporal resolution due to inherent data complexities, including low spatial resolution, limited data availability, and the need for robust feature extraction and classification methods~\cite{huang2020intelligent}.

Deep Learning (DL) has become a promising approach for EEG signal processing, enabling automatic feature extraction from minimally preprocessed data~\cite{wang2020linking}, in contrast to traditional methods reliant on manual feature engineering. However, the effectiveness of DL in BCIs is often limited by the need for large, high-quality training dataset. Acquiring sufficient EEG data is challenging due to the labor-intensive nature of data collection and susceptibility of EEG signals to noise and artifacts~\cite{dimigen2020optimizing}. Generative models have emerged as a potential solution to these challenges. While Generative Adversarial Networks (GANs) have shown promise in generating realistic data, they often face issues like mode collapse and accurately mimicking complexities of EEG signals remains challenging.

Denoising Diffusion Probabilistic Models (DDPMs) offer a compelling alternative for EEG data augmentation, providing a more stable training process and have demonstrated strong performance in generating high-quality images and audio~\cite{muller2022diffusion}. Compared to traditional methods that often introduce artifacts or fail to capture the complexity of EEG patterns through time and frequency-domain transformations, DDPMs generate high-quality, diverse, and realistic synthetic EEG data. Furthermore, unlike parametric models used in earlier techniques, which may not accurately represent the intricacies of EEG data, DDPMs provide a more flexible and robust solution for dataset augmentation, often outperforming traditional methods. DDPMs have shown promise in generating synthetic data.

A common limitation of DDPMs is their tendency to generate highly similar data samples, especially when trained on small datasets. To address this issue, this paper proposes a simple yet effective augmentation technique for DDPMs. By adding noise to the original data before giving input to the DDPMs, the diversity and realism of the generated synthetic EEG data can be significantly enhanced. This augmentation strategy not only improves the effectiveness of data augmentation for EEG research but also leads to substantial improvements in accuracy compared to simply adding noise to the original data without using DDPMs. By increasing the diversity and realism of the generated data, the proposed method can contribute to improved outcomes in various EEG-related applications. Contributions of this paper are as follows:
\begin{itemize}
    \item This study employs a Conditional Denoising Diffusion Model to generate raw synthetic EEG data, representing one of the first attempts in this domain during the time of this work, providing a viable alternative to real datasets.
    \item The diffusion model is trained with noise augmentation from a standard normal distribution, producing synthetic samples that capture meaningful variations of real EEG data rather than exact replicas.
    \item By integrating synthetic data into the training process, the classification accuracy of real EEG data is enhanced, with performance generally improving as the proportion of synthetic data increases.
    \item The efficacy of the proposed method is thoroughly evaluated on the DEAP and SADT datasets, with both raw and synthetic EEG data, establishing its applicability and effectiveness in EEG data classification.
\end{itemize}

The rest of the paper is organized as follows. \Cref{sec_related} provides a comprehensive overview of existing research on emotion classification and synthetic EEG data generation.~\Cref{sec_methodology} describes the methodology used for the whole experiment, including the dataset, diffusion model, and classifiers.~\Cref{sec_results} reports the results, and the work is concluded in~\Cref{sec_conclusion}.


\section{Related Work}
\label{sec_related}

This section provides a comprehensive overview of recent advancements in EEG-based emotion recognition and Synthetic Data Generation (SDG), with a particular emphasis on the contributions of machine and deep learning techniques. The section is divided into two subsections: an analysis of contemporary EEG studies focused on emotion recognition and an exploration of how SDG can enhance these studies. The objective is to summarize the field's current state and identify potential avenues for future research. 

\subsection{Emotion Recognition} 

EEG-based emotion recognition remains a challenging task due to factors such as temporal asymmetry, signal instability, and inter-individual brain variability~\cite{sarma2022emotion}. However, recent advancements in DL have significantly improved emotion recognition techniques. The integration of Long Short-Term Memory (LSTM) models has been instrumental in enhancing EEG signal analysis~\cite{tawhid2022convolutional}. LSTMs effectively capture temporal dynamics and extract relevant features for emotion recognition~\cite{wang2022spatial}. A notable approach involves augmenting LSTM with a multi-view dynamic emotion graph, which considers both electrode channel relationships and temporal information~\cite{xu2023lstm}.

The effectiveness of EEG-based emotion recognition is also contingent on the selection of discriminative features. For instance, one study demonstrated the extraction of narrowband rhythmic components from multichannel EEG, followed by the computation of short-time entropy and energy features for each component. This process was complemented by spatial filtering, with subsequent emotion recognition achieved using a Support Vector Machine (SVM)~\cite{farhana2023emotion}. Another study employed Principal Component Analysis (PCA) for dimensionality reduction before SVM classification, further highlighting the effectiveness of SVM in EEG-based emotion recognition~\cite{doma2020comparative}.

Attention-based methods have emerged as a powerful tool for enhancing neural network performance in EEG analysis by focusing on relevant input segments, drawing inspiration from fields such as psychology, neuroscience, and machine learning~\cite{siddhad2024efficacy}. Approaches include multi-scale feature fusion that incorporates high-level features at different scales~\cite{jiang2023emotion}, pre-trained convolution capsule networks combined with attention mechanism~\cite{liu2023eeg}, and multidimensional representations with global attention~\cite{xiao20224d, xu2023amdet}. These methods demonstrate the potential of attention mechanism in improving accuracy and generalizability in EEG-based emotion recognition.

Hybrid CNN-RNN models have shown promise in EEG emotion recognition. Approaches include cascading CNN features~\cite{nam2021cascaded}, utilizing univariate and multivariate convolution layers to process multichannel EEG~\cite{chao2020emotion}, and employing RNNs for deep feature extraction for domain-invariant representations~\cite{li2020novel}. Additionally, 4D CRNNs~\cite{shen2020eeg} transform differential entropy features into 4D structures and combine CNN and LSTM units. EEGNet~\cite{lawhern2018eegnet} and TSception~\cite{ding2022tsception} offer effective solutions. TSception, a multi-scale CNN, captures temporal and spatial dynamics, while EEGNet focuses on discriminative representation learning. Both frameworks outperform traditional methods.

Graph-based methods, particularly Graph Neural Networks (GNNs), have shown significant potential in enhancing EEG emotion recognition~\cite{zeng2022siam}. By effectively capturing both spatial and temporal EEG features, GNNs can extract complex relationships between EEG channels, leading to improved classification accuracy compared to traditional methods. Recent studies have demonstrated the effectiveness of GNNs in various EEG emotion recognition tasks, including utilizing Fusion Graph Convolutional Networks (FGCNs)~\cite{li2023emotion} and combining GNNs with other models like 1D CNNs~\cite{kim2022eeg}. These advancements underscore the value of graph-based approaches for understanding brain dynamics and improving emotion recognition performance.

\subsection{Synthetic EEG Data} 


Recent advancements in Generative Modeling and Synthetic Data Generation (SDG) have significantly impacted the field of EEG analysis, particularly through the use of Diffusion Models (DMs) to address challenges of data scarcity. For data augmentation, diffusion models address the difficulty of obtaining large-scale, high-quality EEG signals. Zhao et al.~\cite{zhao2024eeg} utilized a diffusion model based on denoising diffusion implicit sampling (DDIM) structure with a low-density sample method to generate emotional EEG signals, which significantly improved emotion recognition accuracy across multiple classifiers. Furthermore, Torma and Szegletes~\cite{torma2025generative} explored enhanced Diffusion Probabilistic Models (DPMs) and efficient sampling techniques, such as DDIMs and progressive distillation, to generate visual evoked potentials and motor imagery signals. Evaluation confirmed that DPMs effectively captured the spatiotemporal characteristics of brain signals, leading to improved performance across various classification models (e.g., ChronoNet~\cite{roy2019chrononet} and EEGNet~\cite{lawhern2018eegnet}) in cross-subject data augmentation experiments. These studies collectively confirm the promising capability of DMs to provide efficient and generalizable EEG data augmentation and robust decoding for deep-learning models.

Prior research on SDG for EEG has explored various non-DM methods. Deep Convolutional Generative Adversarial Networks (DCGANs) have been employed for synthetic EEG generation in epileptic seizure prediction~\cite{rasheed2021generative}, while GANs generally show promise in augmenting EEG data and improving model performance~\cite{dissanayake2022generalized}. Other approaches include the use of Wavelet CNNs and the Extreme Learning Machine Wavelet Auto Encoder (ELM-W-AE) for emotion recognition and dataset augmentation~\cite{ari2022wavelet}. While these methods have demonstrated potential, challenges remain in generating raw EEG data—as opposed to features or image representations—including establishing reliable evaluation metrics~\cite{sun17survey} and ensuring synthetic data maintains enough variation to avoid simple replication of the training set~\cite{aznan2019simulating}.

The application of SDG extends beyond EEG into other biomedical domains. Bt-GAN~\cite{ramachandranpillai2024bt} focuses on generating fair synthetic Electronic Health Record (EHR) data to mitigate biases. While GANs are widely used in SDG, their limitations, such as mode collapse, have led to hybrid approaches like GAN-ST~\cite{rather2024generative}. DPMs have also been compared against GANs for synthetic ECG generation~\cite{adib2023synthetic}, and frameworks like MultiDiffusion~\cite{bar2023multidiffusion} offer flexibility for adapting pre-trained DPMs to various tasks. The computational intensity of standard DPMs has prompted research into more efficient implementations, such as using optical computing~\cite{oguz2024optical}.

This work builds upon the foundation of DDPMs, specifically exploring their application for generating **raw synthetic EEG data** for emotion recognition, a task where traditional methods have often focused only on generating EEG features or their image representations.


\section{Methodology}
\label{sec_methodology}

This study introduces a novel methodology for EEG data augmentation using DDPMs. A key component of our approach is a tailored augmentation module designed specifically to address the unique characteristics of EEG data. By incorporating this module, synthetic EEG data is generated that is more realistic and diverse than traditional methods. The overall methodology involves training a diffusion model on real EEG data, using both real and synthetic data to train a classifier, and evaluating the classifier's performance on a held-out test set of real data. The diffusion process is visually depicted in~\cref{fig_diffusionprocess}. The methodology is divided into two primary components: the diffusion model and the classifiers. The diffusion model component encompasses five subsections: the Gaussian diffusion process, optimizing the denoising model, inference via iterative refinement based on Saharia et al.'s approach~\cite{saharia2022image}, model architecture and noise scheduling, and finally, the proposed augmentation module. The tailored augmentation module is introduced here to encourage the diffusion model to generate diverse EEG samples.


\subsection{Conditional Denoising Diffusion Model}

DDPMs, initially introduced in 2015~\cite{sohl2015deep} and subsequently popularized by Ho et al. in 2020~\cite{ho2020denoising}, have demonstrated remarkable capabilities in image generation. DDPMs employ forward and reverse diffusion processes; the former progressively adds noise to a clean image, while the latter denoises a noisy image sequence to recover the original image. This study adapts the conditional denoising diffusion model proposed by Saharia et al.~\cite{saharia2022image} for image super-resolution to generate conditional and synthetic signals. By employing a U-Net architecture~\cite{ronneberger2015u} with a denoising objective, the model effectively transforms a normal distribution into an empirical data distribution through a series of refinement steps analogous to Langevin dynamics.

The dataset, denoted as $\mathcal{D}=\left\{\boldsymbol{x}_i, \boldsymbol{y}_i\right\}_{i=1}^N$, comprises input-output signal pairs sampled from an unknown distribution $p(\boldsymbol{x}, \boldsymbol{y})$. The DDPM~\cite{ho2020denoising} is adapted for conditional signal generation. Starting from a pure noise sample $\boldsymbol{y}_T \sim \mathcal{N}(\mathbf{0}, \boldsymbol{I})$, a target sample $\boldsymbol{y}_0$ is generated by the conditional DDPM through $T$ refinement stages. The model iteratively refines the output sample based on learned conditional distributions $p_\theta\left(\boldsymbol{y}_{t-1} \mid \boldsymbol{y}_t, \boldsymbol{x}\right)$, resulting in a sequence $\left(\boldsymbol{y}_{T-1}, \boldsymbol{y}_{T-2}, \ldots, \boldsymbol{y}_0\right)$ where the final output $\boldsymbol{y}_0$ is distributed according to $p(\boldsymbol{y} \mid \boldsymbol{x})$, as illustrated in~\cref{fig_diffusionprocess}. The iterative refinement process involves a forward diffusion step that adds Gaussian noise to the output via a fixed Markov chain $q\left(\boldsymbol{y}_t \mid \boldsymbol{y}_{t-1}\right)$, followed by a reverse process that recovers the signal from the noise, conditioned on $\boldsymbol{x}$. The denoising model $f_\theta$ learns this reverse chain by utilizing both the source and noisy target samples to estimate the noise.


\begin{figure}[!t]
    \centering
    \includegraphics[width=\linewidth]{diffusionprocess.pdf}
    \caption{The diffusion model's training process includes forward diffusion $q$ (left to right), where Gaussian noise from a standard normal distribution is added to the real signal $\boldsymbol{x}$ in incremental steps, resulting in $\boldsymbol{x}_t$. Data generation utilizes reverse diffusion $p$ (right to left), where a denoising UNet gradually removes noise from $\boldsymbol{x}_T$, conditioned on $\boldsymbol{x}_\Delta$, to produce the generated signal $\boldsymbol{y}$. Notably, the denoising UNet accepts two inputs: the conditioning information $\boldsymbol{x}_\Delta$ and the noisy signal to be denoised, $\boldsymbol{x}_T$. Here, the Augmentation module controls the noise added to $\boldsymbol{x}$ to generate the conditioning signal $\boldsymbol{x}_{\Delta}$.}
    \label{fig_diffusionprocess}
\end{figure}

\subsubsection{Gaussian Diffusion Process}

Following \cite{ho2020denoising}, the forward diffusion process $q$ gradually adds Gaussian noise to target sample $\boldsymbol{y}_0$ over $T$ iterations, as defined by
\begin{equation}
    q\left(\boldsymbol{y}_{1:T} \mid \boldsymbol{y}_0\right) = \prod\nolimits_{t=1}^T q\left(\boldsymbol{y}_t \mid \boldsymbol{y}_{t-1}\right),
\end{equation}
\begin{equation}
    q\left(\boldsymbol{y}_t \mid \boldsymbol{y}_{t-1}\right) = \mathcal{N}\left(\boldsymbol{y}_t \mid \sqrt{\alpha_t} \boldsymbol{y}_{t-1},\left(1-\alpha_t\right) \boldsymbol{I}\right),
\end{equation}
where $\alpha_{1:t}$ are hyper-parameters such that $0<\alpha_t<1$, determining the noise variance. The variance of the random variables is attenuated by $\sqrt{\alpha_t}$ to ensure boundedness as $t\rightarrow\infty$. The distribution of $\boldsymbol{y}_t$ given $\boldsymbol{y}_0$ is expressed as
\begin{equation}
    \label{eq_ddpm_intermediate}
    q\left(\boldsymbol{y}_t \mid \boldsymbol{y}_0\right)=\mathcal{N}\left(\boldsymbol{y}_t \mid \sqrt{\gamma_t} \boldsymbol{y}_0,\left(1-\gamma_t\right) \boldsymbol{I}\right),
\end{equation}
where $\gamma_t=\prod_{i=1}^t \alpha_i$. The posterior distribution of $\boldsymbol{y}_{t-1}$ given $\left(\boldsymbol{y}_0, \boldsymbol{y}_t\right)$ is given by
\begin{gather}
    \label{eq_posterior}
    q\left(\boldsymbol{y}_{t-1} \mid \boldsymbol{y}_0, \boldsymbol{y}_t\right) = \mathcal{N}\left(\boldsymbol{y}_{t-1} \mid \boldsymbol{\mu}, \sigma^2 \boldsymbol{I}\right), \\
    \boldsymbol{\mu} = \frac{\sqrt{\gamma_{t-1}}\left(1-\alpha_t\right)}{1-\gamma_t} \boldsymbol{y}_0+\frac{\sqrt{\alpha_t}\left(1-\gamma_{t-1}\right)}{1-\gamma_t} \boldsymbol{y}_t,\\
    \sigma^2 = \frac{\left(1-\gamma_{t-1}\right)\left(1-\alpha_t\right)}{1-\gamma_t}.
\end{gather}


\subsubsection{Optimizing the Denoising Model}

The denoising network, crucial for inference in DDPMs presented in~\cref{sec_inferiter}, is conditioned on source sample $\boldsymbol{x}$. The neural denoising model $f_\theta$ is trained to reconstruct the noiseless target sample $\boldsymbol{y}_0$ using $\boldsymbol{x}$ and noisy target signal $\widetilde{\boldsymbol{y}}$. $\widetilde{\boldsymbol{y}}$ is defined as
\begin{equation}
    \label{eq_optyhat}
    \widetilde{\boldsymbol{y}}=\sqrt{\gamma} \boldsymbol{y}_0+\sqrt{1-\gamma} \boldsymbol{\epsilon}, \quad \boldsymbol{\epsilon} \sim \mathcal{N}(\mathbf{0}, \boldsymbol{I}).
\end{equation}
This definition aligns the noisy target sample $\widetilde{\boldsymbol{y}}$ with the marginal distribution at different time steps in the forward diffusion process given in~\cref{eq_ddpm_intermediate}. The model $f_\theta(\boldsymbol{x}, \widetilde{\boldsymbol{y}}, \gamma)$ predicts the noise vector $\boldsymbol{\epsilon}$, conditioned on $\gamma$, similar to generative models~\cite{song2019generative, chen2020wavegrad}. The training objective for $f_\theta$ is given by
\begin{equation}
    \label{eqn:training_objective}
    {E}_{(\boldsymbol{x}, \boldsymbol{y})} {E}_{\boldsymbol{\epsilon}, \gamma}\left\|f_\theta(\boldsymbol{x}, \underbrace{\sqrt{\gamma} \boldsymbol{y}_0+\sqrt{1-\gamma} \boldsymbol{\epsilon}}_{\widetilde{\boldsymbol{y}}}, \gamma)-\boldsymbol{\epsilon}\right\|_p^p,
\end{equation}
where $\boldsymbol{\epsilon} \sim \mathcal{N}(\mathbf{0}, \boldsymbol{I}),(\boldsymbol{x}, \boldsymbol{y})$ is sampled from training data, $p \in \{1,2\}$, and $\gamma \sim p(\gamma)$. The distribution of $\gamma$ significantly impacts model quality and outputs, as detailed in~\Cref{sec_ddpm_arch_noise_schedule}.


\subsubsection{Inference via Iterative Refinement}
\label{sec_inferiter}

The inference process is defined as a reverse Markovian process (similar to~\cite{saharia2022image}), starting from Gaussian noise $\boldsymbol{y}_T$
\begin{align}
    \label{eq_isotropicgaussdist}
    p_\theta\left(\boldsymbol{y}_{0: T} \mid \boldsymbol{x}\right) &= p\left(\boldsymbol{y}_T\right) \prod\nolimits_{t=1}^T p_\theta\left(\boldsymbol{y}_{t-1} \mid \boldsymbol{y}_t, \boldsymbol{x}\right),\\
    \label{eq_isotropicgaussdist2}
    p\left(\boldsymbol{y}_T\right) &= \mathcal{N}\left(\boldsymbol{y}_T \mid \mathbf{0}, \boldsymbol{I}\right),\\
    \label{eq_isotropicgaussdist3}
    p_\theta\left(\boldsymbol{y}_{t-1} \mid \boldsymbol{y}_t, \boldsymbol{x}\right) &= \mathcal{N}\left(\boldsymbol{y}_{t-1} \mid \mu_\theta\left(\boldsymbol{x}, \boldsymbol{y}_t, \gamma_t\right), \sigma_t^2 \boldsymbol{I}\right).
\end{align}

The inference process is defined in terms of isotropic Gaussian conditional distributions, $p_\theta\left(\boldsymbol{y}_{t-1} \mid \boldsymbol{y}_t, \boldsymbol{x}\right)$. When the forward process noise variance $\alpha_{1: T} \approx 1$, the optimal reverse process $p\left(\boldsymbol{y}_{t-1} \mid \boldsymbol{y}_t, \boldsymbol{x}\right)$ is approximately Gaussian~\cite{sohl2015deep}. Accordingly, the choice of Gaussian conditionals in the inference process (\ref{eq_isotropicgaussdist3}) provides a reasonable fit to the true reverse process. Moreover, for the process to start from pure Gaussian noise, $1-\gamma_T$ needs to be sufficiently large (i.e., close to one) from~\cref{eqn:training_objective}, aligning with the prior in~\cref{eq_isotropicgaussdist2}, allowing the sampling process to start at pure Gaussian noise. 

The denoising model $f_\theta$ estimates $\boldsymbol{\epsilon}$ from noisy samples $\widetilde{\boldsymbol{y}}$ including $\boldsymbol{y}_t$. Thus, approximating $\boldsymbol{y}_0$ by reconfiguring~\cref{eq_optyhat}:
\begin{equation}
    \hat{\boldsymbol{y}}_0=\frac{1}{\sqrt{\gamma_t}}\left(\boldsymbol{y}_t-\sqrt{1-\gamma_t} f_\theta\left(\boldsymbol{x}, \boldsymbol{y}_t, \gamma_t\right)\right).
\end{equation}
Following \cite{ho2020denoising}, $\hat{\boldsymbol{y}}_0$ is substituted into the posterior distribution of $q\left(\boldsymbol{y}_{t-1} \mid \boldsymbol{y}_0, \boldsymbol{y}_t\right)$ in~\cref{eq_posterior}, to parameterize the mean of $p_\theta\left(\boldsymbol{y}_{t-1} \mid \boldsymbol{y}_t, \boldsymbol{x}\right)$ as
\begin{equation}
    \boldsymbol{\mu}_\theta\left(\boldsymbol{x}, \boldsymbol{y}_t, \gamma_t\right)=\frac{1}{\sqrt{\alpha_t}}\left(\boldsymbol{y}_t-\frac{1-\alpha_t}{\sqrt{1-\gamma_t}} f_\theta\left(\boldsymbol{x}, \boldsymbol{y}_t, \gamma_t\right)\right),
\end{equation}
and the variance of $p_\theta\left(\boldsymbol{y}_{t-1} \mid \boldsymbol{y}_t, \boldsymbol{x}\right)$ is set to ($1-\alpha_t$), in the forward process~\cite{ho2020denoising}. Thus, each iterative refinement step in the model is structured as
\begin{equation}
    \boldsymbol{y}_{t-1} \leftarrow \frac{1}{\sqrt{\alpha_t}}\left(\boldsymbol{y}_t-\frac{1-\alpha_t}{\sqrt{1-\gamma_t}} f_\theta\left(\boldsymbol{x}, \boldsymbol{y}_t, \gamma_t\right)\right)+\sqrt{1-\alpha_t} \boldsymbol{\epsilon}_t,
\end{equation}
where $\boldsymbol{\epsilon}_t \sim \mathcal{N}(\mathbf{0}, \boldsymbol{I})$. This resembles the Langevin dynamics step with $f_\theta$ providing an estimate of the gradient of the data log density. 


\subsubsection{Model Architecture and Noise Schedulers}
\label{sec_ddpm_arch_noise_schedule}

The model adopts a U-Net architecture similar to the one in DDPM~\cite{ho2020denoising}, integrating self-attention and modifications from \cite{song2020score}. Specifically, it replaces DDPM's original residual blocks with those from BigGAN and rescales skip connections by $\frac{1}{\sqrt{2}}$. The number of residual blocks and channel multipliers at various resolutions is increased to enhance the model. The model is conditioned on input $\boldsymbol{x}$ by introducing noise to the source sample and concatenating it with $\boldsymbol{y}_t$ along the channel dimension. The training noise schedule follows \cite{chen2020wavegrad}, employing a piece-wise distribution for $\gamma$, $p(\gamma)=\sum_{t=1}^T \frac{1}{T} U\left(\gamma_{t-1}, \gamma_t\right)$. Training involves uniformly sampling a time step $t \sim\{0, \ldots, T\}$ and subsequently sampling $\gamma \sim U\left(\gamma_{t-1}, \gamma_t\right)$.

\begin{figure}[!t]
    \centering
    \includegraphics[width=\linewidth]{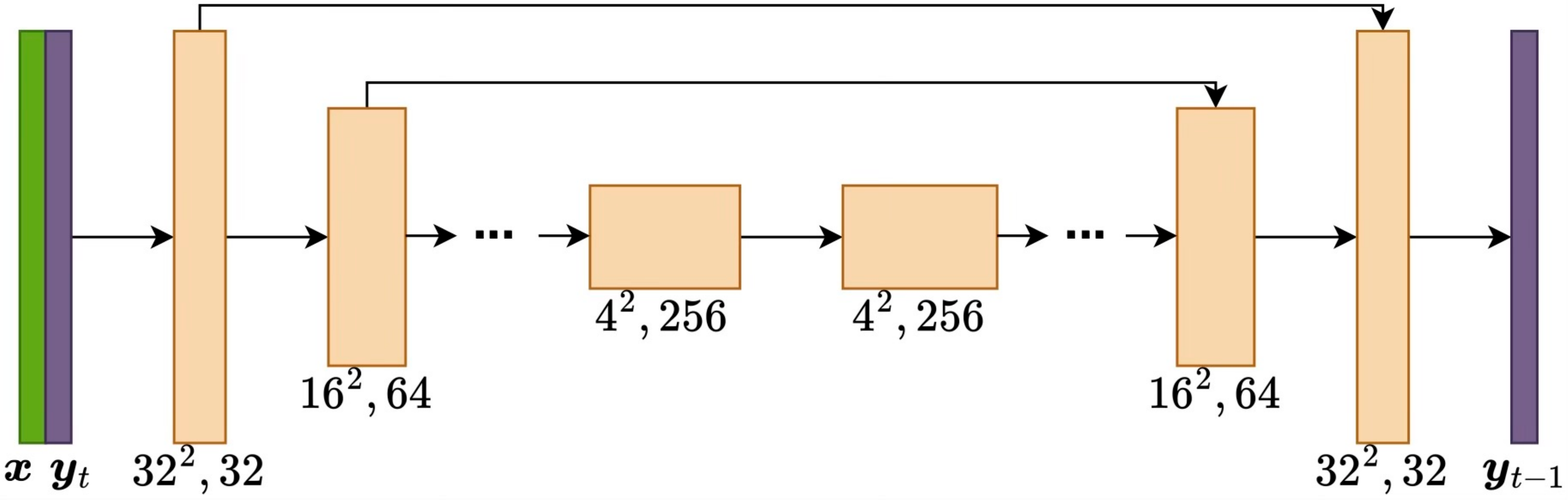}
    \caption{U-Net architecture of diffusion model. The source sample $\boldsymbol{x}$ is concatenated with the target sample $\boldsymbol{y}_t$. Self-attention is performed on 16$\times$16 feature maps.}
    \label{fig_unet_arch}
\end{figure}

This study adopts a technique from \cite{chen2020wavegrad}, which efficiently generates samples by directly conditioning on $\gamma$ rather than on $t$ as in \cite{ho2020denoising}, allowing flexibility in choosing diffusion steps and noise schedules. Assuming a linear noise schedule, the maximum diffusion steps is set to 100. An inexpensive hyper-parameter search over start and end noise levels was performed, avoiding the need for model retraining~\cite{chen2020wavegrad}. The diffusion model employs a denoising U-Net (\cref{fig_unet_arch}) to process noisy target and source conditioning samples.


\subsubsection{Augmentation Module}

The diffusion process involves two samples: a source sample $\boldsymbol{x}$ and a target sample $\boldsymbol{y}$. To modify the source sample, an augmentation module is employed that adds noise from a standard normal distribution $Z$, creating a condition sample as shown in~\cref{fig_diffusionprocess}. This is formalized by:
\begin{equation}
    \boldsymbol{x}_\Delta = \boldsymbol{x} + \Delta \times Z,
\end{equation}
where $\Delta$ exhibits variability. These variations are crucial for the diffusion model as they lead to the sample generation influenced by noise. While the model relies on real samples as a foundation for learning, it adapts to produce new samples with subtle differences.


\subsection{Classifiers}



This research employs a comparative evaluation of several advanced classifiers, for EEG-based emotion recognition and vigilance detection. The classifiers selected for this analysis are SVM~\cite{cortes1995support}, EEGNet~\cite{lawhern2018eegnet}, TSception~\cite{ding2022tsception}, and LMDA-Net~\cite{miao2023lmda}. 

SVM, a well-established machine learning algorithm, is utilized for its capacity to effectively differentiate between classes by optimizing the separation margin. EEGNet, which employs a CNN architecture, is incorporated due to its proficiency in extracting spatio-temporal features from EEG signals through depth-wise and separable convolutions. TSception, a more contemporary method, is included as it integrates dynamic temporal and asymmetric spatial layers, enabling the capture of intricate EEG patterns that correlate with emotional states. While TSception holds a central role as the primary classifier in this study, the inclusion of SVM and EEGNet facilitates a thorough assessment of the influence of synthetic data on classification outcomes. Furthermore, LMDA-Net, a lightweight multi-dimensional attention network, was added to the analysis to broaden the scope and include newer methodologies, specifically when examining the SADT dataset. This expanded selection of classifiers ensures a robust and comprehensive evaluation of the synthetic data's impact on EEG-based emotion recognition.

\begin{table*}[!t]
	\centering    
    \caption{Comparison of SVM, EEGNet, and TSception accuracy on the DEAP dataset. Performance is shown for real-only data vs. Real + Synthetic data (1:1 ratio), generated using GAN and our method at noise levels $\Delta = 0.01, 0.05, 0.1$, vanilla conditional diffusion ($\Delta = 0$). Accuracies are presented with 95\% confidence intervals.}
	\resizebox{\linewidth}{!}{
    \begin{tabular}{l c c c c c c c c c c c}
		\toprule
        & \multirow{2}{*}{\textbf{Baseline}} & \multicolumn{2}{c}{\textbf{Vanilla Diffusion}} & \multicolumn{2}{c}{\textbf{Proposed Method}} & \multicolumn{2}{c}{\textbf{Proposed Method}} & \multicolumn{2}{c}{\textbf{Proposed Method}} & \multicolumn{2}{c}{\multirow{2}{*}{\textbf{GAN}~\cite{goodfellow2020generative}}} \\
        
        & & \multicolumn{2}{c}{\textbf{($\Delta=0$)}} & \multicolumn{2}{c}{\textbf{($\Delta=0.01$)}} & \multicolumn{2}{c}{\textbf{($\Delta=0.05$)}} & \multicolumn{2}{c}{\textbf{($\Delta=0.1$)}} & \\ 
        
        
        & \textbf{Real} & \multicolumn{2}{c}{\textbf{Real + Synthetic}} & \multicolumn{2}{c}{\textbf{Real + Synthetic}} & \multicolumn{2}{c}{\textbf{Real + Synthetic}} & \multicolumn{2}{c}{\textbf{Real + Synthetic}} & \multicolumn{2}{c}{\textbf{Real + Synthetic}} \\

        \midrule
        \textbf{Arousal} & \textbf{Accuracy} & \textbf{Accuracy} & \textbf{Gain} & \textbf{Accuracy} & \textbf{Gain} & \textbf{Accuracy} & \textbf{Gain} & \textbf{Accuracy} & \textbf{Gain} & \textbf{Accuracy} & \textbf{Gain} \\
        \midrule

        SVM~\cite{cortes1995support} & $58.71 \pm 0.06$ & $58.71 \pm 0.21$ & $-0.01$ & $59.12 \pm 0.16$ & $0.41$ & $59.14 \pm 0.08$ & $\textbf{0.42}$ & $59.07 \pm 0.09$ & $0.36$ & $59.11 \pm 0.20$ & $0.40$ \\
        EEGNet~\cite{lawhern2018eegnet} & $67.59 \pm 0.23$ & $67.58 \pm 0.32$ & $-0.01$ & $68.24 \pm 0.29$ & $\textbf{0.66}$ & $68.12 \pm 0.32$ & $0.53$ & $67.77 \pm 0.38$ & $0.19$ & $66.76 \pm 0.69$ & $-0.82$ \\
        TSception~\cite{ding2022tsception} & $67.25 \pm 1.01$ & $67.29 \pm 1.05$ & $0.04$ & $69.19 \pm 1.63$ & $\textbf{1.94}$ & $68.24 \pm 0.90$ & $0.99$ & $67.81 \pm 0.84$ & $0.56$ & $66.73 \pm 0.84$ & $-0.52$ \\

        \midrule
        \textbf{Dominance} & \textbf{Accuracy} & \textbf{Accuracy} & \textbf{Gain} & \textbf{Accuracy} & \textbf{Gain} & \textbf{Accuracy} & \textbf{Gain} & \textbf{Accuracy} & \textbf{Gain} & \textbf{Accuracy} & \textbf{Gain} \\
        \midrule

        SVM~\cite{cortes1995support} & $63.02 \pm 0.07$ & $63.07 \pm 0.14$ & $0.05$ & $63.32 \pm 0.06$ & $\textbf{0.31}$ & $63.23 \pm 0.08$ & $0.21$ & $63.25 \pm 0.18$ & $0.24$ & $63.11 \pm 0.06$ & $0.09$ \\
        EEGNet~\cite{lawhern2018eegnet} & $69.53 \pm 0.43$ & $69.56 \pm 0.48$ & $0.02$ & $70.14 \pm 0.54$ & $0.61$ & $69.83 \pm 0.13$ & $0.29$ & $69.92 \pm 0.22$ & $0.39$ & $68.86 \pm 0.49$ & $-0.67$ \\
        TSception~\cite{ding2022tsception} & $69.25 \pm 1.19$ & $69.25 \pm 1.24$ & $0.00$ & $70.19 \pm 0.88$ & $\textbf{0.94}$ & $70.07 \pm 1.10$ & $0.82$ & $69.19 \pm 0.57$ & $-0.06$ & $68.56 \pm 0.48$ & $-0.69$ \\

		\midrule
        \textbf{Liking} & \textbf{Accuracy} & \textbf{Accuracy} & \textbf{Gain} & \textbf{Accuracy} & \textbf{Gain} & \textbf{Accuracy} & \textbf{Gain} & \textbf{Accuracy} & \textbf{Gain} & \textbf{Accuracy} & \textbf{Gain} \\
        \midrule

        SVM~\cite{cortes1995support} & $66.51 \pm 0.09$ & $66.48 \pm 0.11$ & $-0.03$ & $66.61 \pm 0.07$ & $0.10$ & $66.62 \pm 0.07$ & $\textbf{0.11}$ & $66.57 \pm 0.08$ & $0.06$ & $66.57 \pm 0.03$ & $0.06$\\
        EEGNet~\cite{lawhern2018eegnet} & $69.66 \pm 0.42$ & $69.69 \pm 0.40$ & $0.02$ & $70.35 \pm 0.36$ & $\textbf{0.69}$ & $70.23 \pm 0.44$ & $0.57$ & $70.03 \pm 0.27$ & $0.36$ & $68.83 \pm 0.22$ & $-0.83$ \\
        TSception~\cite{ding2022tsception} & $70.15 \pm 0.55$ & $70.09 \pm 0.62$ & $-0.06$ & $71.50 \pm 0.70$ & $\textbf{1.34}$ & $70.99 \pm 0.79$ & $0.84$ & $70.55 \pm 1.00$ & $0.40$ & $69.41 \pm 0.67$ & $-0.74$ \\

		\midrule
        \textbf{Valence} & \textbf{Accuracy} & \textbf{Accuracy} & \textbf{Gain} & \textbf{Accuracy} & \textbf{Gain} & \textbf{Accuracy} & \textbf{Gain} & \textbf{Accuracy} & \textbf{Gain} & \textbf{Accuracy} & \textbf{Gain} \\
        \midrule

        SVM~\cite{cortes1995support} & $56.42 \pm 0.10$ & $56.48 \pm 0.22$ & $0.06$ & $56.66 \pm 0.21$ & $\textbf{0.25}$ & $56.54 \pm 0.12$ & $0.12$ & $56.53 \pm 0.08$ & $0.11$ & $56.55 \pm 0.16$ & $0.13$ \\
        EEGNet~\cite{lawhern2018eegnet} & $66.50 \pm 0.37$ & $66.48 \pm 0.37$ & $-0.02$ & $67.64 \pm 0.55$ & $1.13$ & $67.81 \pm 0.34$ & $\textbf{1.31}$ & $67.42 \pm 0.46$ & $0.92$ & $65.66 \pm 0.68$ & $-0.84$ \\
        TSception~\cite{ding2022tsception} & $65.85 \pm 1.17$ & $65.90 \pm 1.30$ & $0.05$ & $67.64 \pm 0.85$ & $1.79$ & $67.61 \pm 0.27$ & $1.76$ & $67.77 \pm 1.03$ & $\textbf{1.91}$ & $65.67 \pm 0.79$ & $-0.18$ \\
		\bottomrule
	\end{tabular}}
	\label{tab_results}
\end{table*}

\begin{table}[!t]
	\centering    
    \caption{SADT dataset: Classifier accuracy and performance comparison with $\Delta = 0.01$ synthetic data. Results show 95\% confidence intervals at 100\% real/synthetic sample proportion.}
    \begin{tabular}{l c c c}
		\toprule
        & \textbf{Real} & \multicolumn{2}{c}{\textbf{Real + Synthetic}} \\
        \midrule
        \textbf{SADT} & \textbf{Accuracy} & \textbf{Accuracy} & \textbf{Gain} \\
        \midrule
        SVM~\cite{cortes1995support} & $63.26 \pm 0.51$ & $65.08 \pm 0.56$ & $\textbf{1.82}$ \\
        EEGNet~\cite{lawhern2018eegnet} & $85.80 \pm 1.18$ & $86.24 \pm 1.01$ & $\textbf{0.44}$ \\
        TSception~\cite{ding2022tsception} & $84.26 \pm 1.94$ & $84.63 \pm 0.74$ & $\textbf{0.37}$ \\
        LMDA-Net~\cite{miao2023lmda} & $72.18 \pm 2.17$ & $73.44 \pm 1.02$ & $\textbf{1.26}$ \\
		\bottomrule
	\end{tabular}
	\label{tab_results_sadt}
\end{table}


\section{Results and Discussion}
\label{sec_results}

The efficacy of this study is evaluated using real EEG data and synthetic data generated with varying noise levels ($\Delta$). Evaluation is conducted on SVM, EEGNet, and TSception classifiers, with the following key experimental aspects:
\begin{itemize}
    \item Training classifiers on purely real data, followed by a combination of real and synthetic data.
    \item Investigating accuracy changes relative to the number of synthetic EEG samples included in the training.
    \item Assessing classification accuracy variations with synthetic data generated with different noise levels ($\Delta$).
    \item Comparing performance outcomes when using synthetic data generated by vanilla diffusion and GAN.
\end{itemize}
All classifiers undergo testing on a dedicated real data test set.


\subsection{Experimental Data}

This study employed two distinct EEG datasets: the DEAP~\cite{koelstra2011deap} and SADT~\cite{cao2019multi}. The DEAP dataset, a widely used multimodal resource for affective computing, comprises EEG and physiological signals from 32 participants who viewed 40 one-minute music videos. These videos were rated on a 1-9 scale, subsequently binarized into `high' and `low' affective states using a threshold of 5. For the purpose of this study, these binarized labels were used to represent the two-quadrant emotion recognition scheme. While a four-quadrant scheme offers a more nuanced perspective on emotional states, the primary objective of this work was to demonstrate the improved classification performance of various machine learning and deep learning classifiers when augmented with the generated synthetic EEG data, rather than to extensively compare different emotion schemes. Preprocessed EEG data, downsampled to 128 Hz from the original 512 Hz, was utilized, incorporating 32 channels with EOG artifact removal, band-pass filtering, common average referencing, and Geneva convention reordering. Following a 3-second pre-trial baseline removal, the data was segmented into 60-second trials and further epoched into one-second intervals, resulting in samples of shape (1, 32, 128).

SADT dataset, focused on alertness and drowsiness, consisted of EEG recordings from 11 subjects. It contained 2022 epochs sampled at 128 Hz across 30 channels. These epochs were labeled as alert or drowsy. To address the difference in channel count (DEAP uses 32 channels, while SADT uses 30), a specific strategy was employed for the transfer learning process. The last two channels of the SADT dataset were appended as channels 31 and 32. This successfully leveraged the 32-channel model trained on the DEAP dataset for transfer learning on SADT, maintaining consistency in the input dimensions. Both datasets were partitioned into training, validation, and testing sets using a 70:15:15 ratio adopting a subject-independent scheme, with stratification where applicable, and reshaped consistently for compatibility with the proposed model. The inclusion of balanced accuracy and subject-specific data partitioning in our evaluation methodology ensures a more rigorous and transparent assessment of our model's performance.


\subsection{Implementation Details}

The experimental setup involved a DELL Precision 7820 Tower Workstation, with Ubuntu 22.04 OS, Intel Core(TM) Xeon Silver 4216 CPU, and NVIDIA RTX A2000 12GB GPU. This facilitated the implementation of DL models using Python 3.10 and the PyTorch library. The Adam optimizer, known for its computational efficiency, was used with default parameters ($\eta$ = 0.001, $\beta_1$ = 0.9, $\beta_2$ = 0.999), and a linear warmup was applied over 10,000 training steps. EEGNet and TSception were trained for 100 epochs, with batches of 16 and a learning rate of $1e-4$. For SVM, the Radial Basis Function (RBF) kernel from scikit-learn~\cite{sklearn} was used with default settings. Classification accuracy was determined through stratified five-fold cross-validation and averaging the results.

For diffusion experiments, $T$ = 500 was maintained and the $\gamma_t$ values were uniformly spaced. While larger $T$ values can potentially improve model performance, using $T$ = 2000 as in \cite{saharia2022image} results in noiseless samples, contrary to the goals of this study. Additionally, this setting also accelerates sample generation (or inference) compared to earlier models~\cite{ho2020denoising, song2020score}, which required between 1000 and 2000 diffusion steps. The diffusion model training involved batches of 32 and a 0.2 dropout rate for 1 million steps, utilizing the latest checkpoint. Each training epoch took an average of 785.89 seconds.


\subsection{Synthetic Data Generation and Validation}

\begin{figure}[!t]
    \centering
    \includegraphics[width=\linewidth]{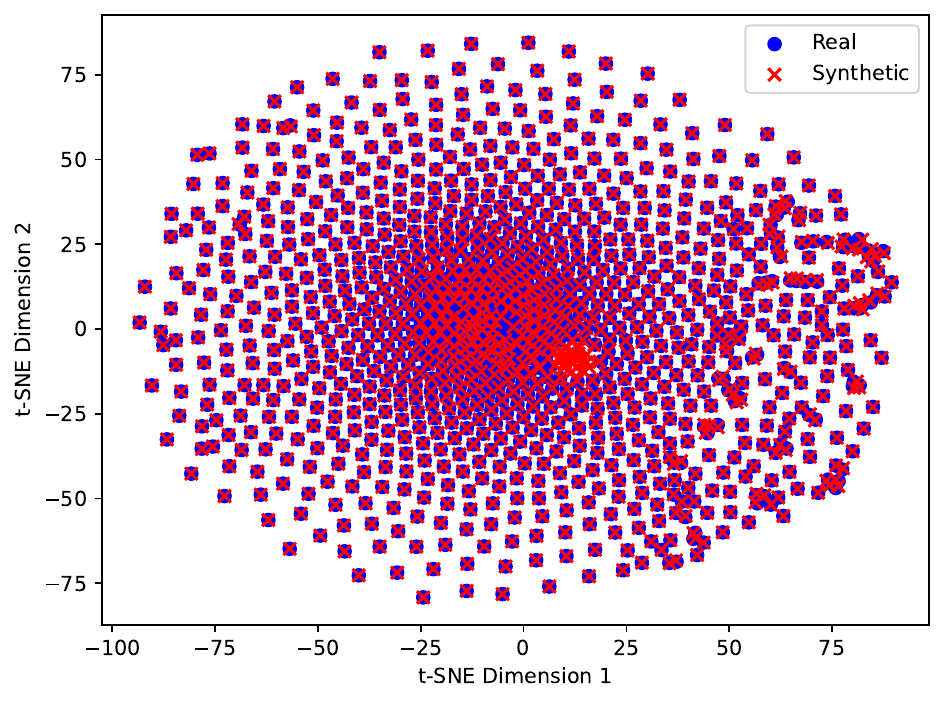}
    \caption{t-SNE visualization comparing real and synthetic EEG data. The spatial closeness of clusters indicates the degree of similarity between real and synthetic samples.}
    \label{fig_tsne}
\end{figure}

To thoroughly assess the quality and fidelity of generated synthetic EEG data, both quantitative visualization and qualitative inspection methods were employed. \cref{fig_tsne} presents a two-dimensional visualization generated using t-Distributed Stochastic Neighbor Embedding (t-SNE)~\cite{van2008visualizing}, a powerful dimensionality reduction technique that is particularly effective at visualizing high-dimensional data while preserving local structures. In this visualization, the intrinsic distributions of both real and synthetic EEG data samples are compared. The spatial proximity of clusters within the t-SNE plot directly reflects the similarity between the data distributions. Specifically, if the clusters representing the real and synthetic samples are highly overlapping or closely aligned, it indicates a high degree of fidelity, suggesting that the synthetic data effectively approximates the underlying distribution of the real dataset. Results in \cref{fig_tsne} demonstrate significant overlap, indicating a strong resemblance, providing a quantitative measure of the similarity in data distributions.

\begin{figure}[ht] 
  \begin{subfigure}[b]{\linewidth}
    \centering
    \includegraphics[width=\linewidth]{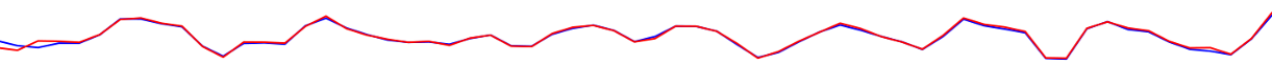}
    \caption{} 
    \label{fig_synsam01} 
  \end{subfigure}\\
  \begin{subfigure}[b]{\linewidth}
    \centering
    \includegraphics[width=\linewidth]{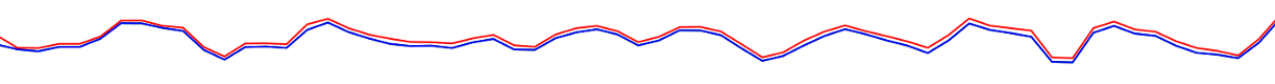}
    \caption{} 
    \label{fig_synsam05} 
  \end{subfigure}\\
  \begin{subfigure}[b]{\linewidth}
    \centering
    \includegraphics[width=\linewidth]{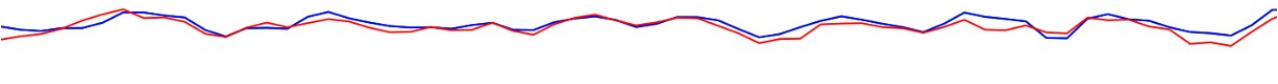}
    \caption{} 
    \label{fig_synsam1} 
  \end{subfigure}
  \caption{Visual Comparison of Real (blue) and Synthetic (red) EEG Samples for $\Delta$: (a) $\Delta=0.01$, (b) $\Delta=0.05$, and (c) $\Delta=0.1$. The high visual similarity in temporal patterns and amplitude dynamics confirms the synthetic data effectively replicates key characteristics of real EEG signals.}
  \label{fig_synsam} 
\end{figure}

Beyond statistical distributions, it is crucial to visually inspect the time-series characteristics of the generated synthetic EEG signals. \cref{fig_synsam} provides a direct visual comparison between real and corresponding synthetic EEG samples for varying values of $\Delta$, a key parameter in proposed generative model. Each subfigure displays a pair of EEG signals: the blue signal represents a real EEG sample, and the red signal depicts a synthetic sample generated by the proposed model. The visual similarity across temporal patterns, amplitude dynamics, and frequency characteristics serves as a strong indicator of the synthetic data's quality. This suggests that the synthetic data effectively captures the complex nuances inherent in real EEG recordings, indicating a remarkable visual consistency, especially for $\Delta=0.01$. This visual validation complements the t-SNE analysis by confirming the preservation of critical signal characteristics at a granular level.


\subsection{Effect of Synthetic Data in Classification Performance}

\subsubsection{DEAP Dataset: Impact of Synthetic Data on Emotion Recognition}

It was hypothesized that classifier performance would be enhanced by synthetic data, generated through the introduction of controlled variations to real samples via augmented noise ($\Delta$). To test this hypothesis, classifiers (SVM, EEGNet, and TSception) were trained on a combination of real and synthetic data, with equal proportions of each, and subsequently evaluated on real data alone. This approach was compared to training solely on real data.

Classification accuracy for each model across different noise levels ($\Delta = 0.01, 0.05, 0.1$) is presented in \cref{tab_results}, along with a baseline where no noise was added ($\Delta=0$, equivalent to replicating real data) and a comparison with GAN-generated data. The baseline accuracy achieved using only real data is represented in the "Real" column. The confirmation that replicating real data does not alter performance is provided by the ``Vanilla Diffusion'' column ($\Delta=0$). All reported accuracy values are presented with their corresponding 95\% confidence intervals, calculated through 5-fold stratified cross-validation.

Across all classifiers, improvements in accuracy were observed with the inclusion of synthetic data from our diffusion model, particularly at lower noise levels ($\Delta = 0.01, 0.05$). Modest gains were shown by SVM, with the highest improvements observed at $\Delta = 0.05$ for Arousal and $\Delta = 0.01$ for Dominance, Liking, and Valence. More substantial improvements were exhibited by EEGNet and TSception, especially at $\Delta = 0.01$, demonstrating the effectiveness of our approach in enhancing deep learning model performance. Notably, a performance drop was experienced by both EEGNet and TSception when trained with GAN-generated data, highlighting the superior quality of our diffusion-based synthetic data. These improvements and drops in performance were statistically significant, as evidenced by the non-overlapping 95\% confidence intervals between the baseline and augmented results.

It is suggested by the consistent improvement at lower noise levels that subtle variations in synthetic data effectively augment the real dataset, enhancing feature learning and generalization. The ability of our synthetic data to enrich complex feature representations is underscored by the higher gains observed with deep learning models, particularly TSception.

To further validate the efficacy of our diffusion-based synthetic data, an experiment was conducted using purely Gaussian noise, unrelated to EEG patterns. As shown in \cref{tab_noise}, a decrease in accuracy for Arousal and Dominance, and negligible changes for Liking and Valence, were observed when training with Gaussian noise. This demonstrates that the improvements observed with our diffusion-generated data are not merely due to data augmentation, but rather to the introduction of meaningful, EEG-like variations that enhance feature learning.

\subsubsection{SADT Dataset: Enhancing Vigilance State Classification}

The investigation was expanded to the SADT dataset to assess the influence of synthetic data on vigilance state classification, specifically distinguishing between alert and drowsy states. Notably, the synthetic samples used for the SADT dataset were generated using the diffusion model previously trained on the DEAP dataset, without training a new model specifically for SADT. This approach allows us to evaluate not only the impact of synthetic data but also demonstrates the potential for transfer learning in synthetic data generation to improve classification accuracy on a new, related dataset. \cref{tab_results_sadt} details the accuracy and performance enhancements achieved by SVM, EEGNet, TSception, and LMDA-Net when trained on a combined dataset of real and synthetic data (generated at a noise level of $\Delta=0.01$), in comparison to training solely on real SADT data.

In alignment with the findings obtained from the DEAP dataset, the incorporation of synthetic data resulted in improved accuracy across all evaluated classifiers. Notably, SVM demonstrated the most substantial improvement, with a 1.82\% increase in accuracy. This suggests that the synthetic data effectively acted as a regularizer, enhancing the model's capacity to differentiate between alert and drowsy states. LMDA-Net also exhibited a notable gain of 1.26\%. EEGNet and TSception showed modest improvements of 0.44\% and 0.37\%, respectively, indicating that the synthetic data contributed positively to their performance. The comparatively smaller gains observed in deep learning models, which already achieved high baseline accuracy, can be attributed to their inherent ability to learn complex features, thereby leaving less room for further enhancement.

The consistent improvements across all models underscore the effectiveness of our diffusion-based synthetic data generation method in enhancing EEG-based classification performance. The ability of our approach to introduce meaningful variations that augment real data and improve model generalization is highlighted by the consistent improvements, particularly at lower noise levels. The quality and relevance of our synthetic EEG data are further validated by the reduced performance observed when using GAN-generated data and Gaussian noise. The 95\% confidence intervals further indicate a level of reliability and consistency in the observed improvements, suggesting that the synthetic data effectively augments the real dataset, leading to more robust and accurate classification performance.

This overall trend, observed across both the DEAP and SADT datasets, highlights the effectiveness of our diffusion-based synthetic data generation method in enhancing EEG-based classification performance. Our approach's ability to introduce meaningful variations that augment real data and improve model generalization is highlighted by the consistent improvements, particularly at lower noise levels.



\begin{table}[!t]
	\centering
	\caption{Impact of training TSception by combining real data and Gaussian noise samples used in equal proportions (i.e., 100\%) for emotion recognition using DEAP dataset.}
	\begin{tabular}{l c c c}
		\toprule
        & \textbf{Real} & \multicolumn{2}{c}{\textbf{Real + Noise}} \\
        \midrule
        \textbf{TSception} & \textbf{Accuracy} & \textbf{Accuracy} & \textbf{Gain} \\
        \midrule
        \textbf{Arousal} & $67.25 \pm 1.01$ & $66.92 \pm 0.69$ & $-0.33$ \\
        \textbf{Dominance} & $69.25 \pm 1.19$ & $68.49 \pm 0.73$ & $-0.76$ \\
        \textbf{Liking} & $70.15 \pm 0.55$ & $70.11 \pm 0.83$ & $-0.05$ \\
        \textbf{Valence} & $65.85 \pm 1.17$ & $65.98 \pm 0.87$ & $0.13$ \\
		\bottomrule
	\end{tabular}
	\label{tab_noise}
\end{table}

\begin{table}[!t]
	\centering
	\caption{Maximum accuracy achieved in ablation experiments by training TSception combining real and synthetic data ($\Delta=0.01$) using DEAP dataset.}
	\begin{tabular}{l c c c}
		\toprule
        & \textbf{Real} & \multicolumn{2}{c}{\textbf{Real + Synthetic}} \\
        \midrule
        \textbf{TSception} & \textbf{Accuracy} & \textbf{Accuracy} & \textbf{Gain} \\
        \midrule
        \textbf{Arousal} & $67.25 \pm 1.01$ & $71.44 \pm 1.06$ & $5.60$ \\
        \textbf{Dominance} & $69.25 \pm 1.19$ & $72.42 \pm 2.01$ & $4.32$ \\
        \textbf{Liking} & $70.15 \pm 0.55$ & $73.12 \pm 1.02$ & $3.90$ \\
        \textbf{Valence} & $65.85 \pm 1.17$ & $70.07 \pm 0.80$ & $4.97$ \\
		\bottomrule
	\end{tabular}
	\label{tab_max}
\end{table}

\begin{figure*}[!t]
    \centering
    \begin{subfigure}[b]{0.55\textwidth}
        \centering
        \includegraphics[width=\textwidth]{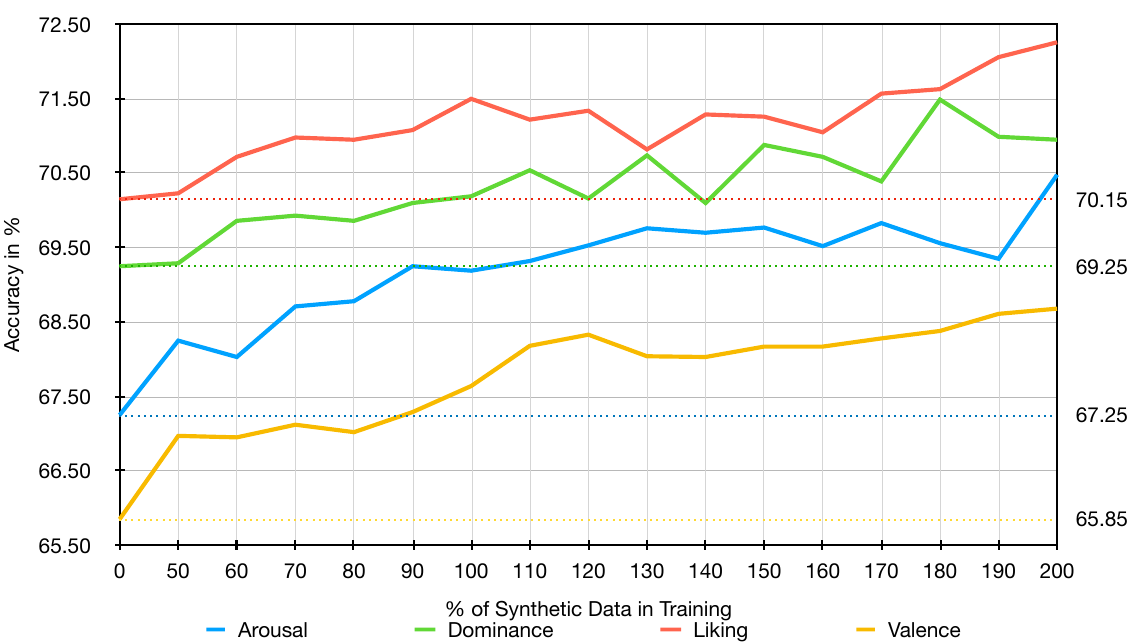}
        \caption{}
        \label{fig_ablation_10}
    \end{subfigure}
    \hfill
    \begin{subfigure}[b]{0.43\textwidth}
        \centering
        \includegraphics[width=\textwidth]{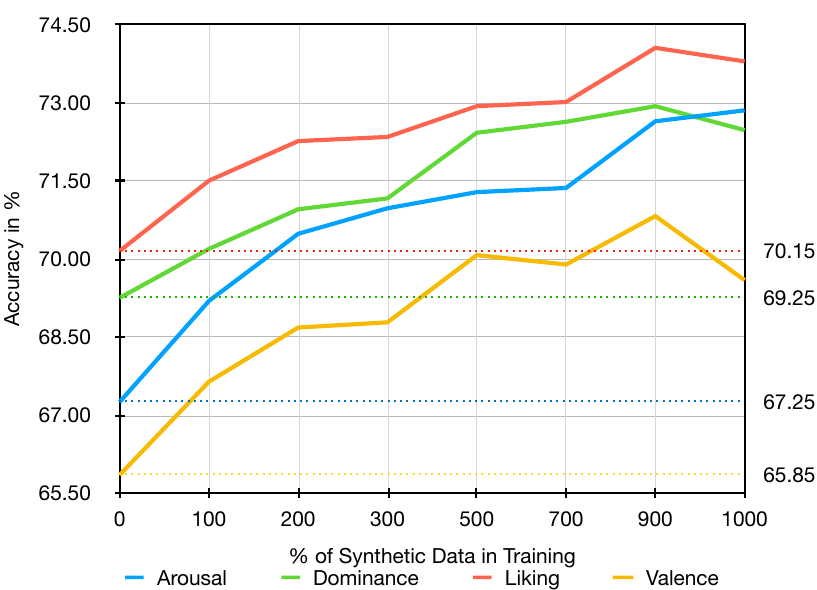}
        \caption{}
        \label{fig_ablation_100}
    \end{subfigure}
    \caption{Charts illustrating classifier accuracy for different emotional states as the proportion of diffusion-generated ($\Delta=0.01$) training data increases. Dotted lines show baseline accuracy (no synthetic data). `100\%' indicates an equal number of synthetic and real samples. Chart (a) shows 10\% increments of synthetic data, while (b) shows 100\% and 200\% increments.}
    \label{fig_ablation}
\end{figure*}

\subsection{Effect of the Number of Synthetic Samples Used in Training}

A study was conducted to investigate the effect of varying synthetic data volumes on classifier performance using DEAP dataset. The study focused on data with $\Delta$ = 0.01, which previously demonstrated superior performance over other noise variants. As shown in~\cref{fig_ablation}, the model's accuracy correlates with the percentage of synthetic data employed during classifier training, with data proportions ranging from 50\% to 1000\%. The x-axis represents the synthetic data percentage, while the y-axis indicates accuracy. The baseline accuracy (dotted lines) serves as a reference to show the relative improvement gained by including synthetic data during classifier training. \cref{tab_max} presents the maximum accuracy achieved by TSception when trained on both real and synthetic EEG data, compared to using only real data. The results demonstrate a consistent improvement in classification accuracy when incorporating synthetic samples. Notably, arousal (blue) exhibits highest accuracy, followed by valence (yellow), dominance (red), and liking (green). The results indicate a correlation between the inclusion of synthetic data and classifier performance, with the highest accuracy achieved at a 900\% synthetic data mix. Beyond this threshold, however, further increases in synthetic data did not lead to noticeable improvements in classification accuracy. These findings suggest that augmenting training data with synthetic EEG signals can effectively enhance emotion recognition capabilities.


\subsection{Interpretability Analysis}

\begin{figure}[ht] 
  \begin{subfigure}[b]{0.5\linewidth}
    \centering
    \includegraphics[width=\linewidth]{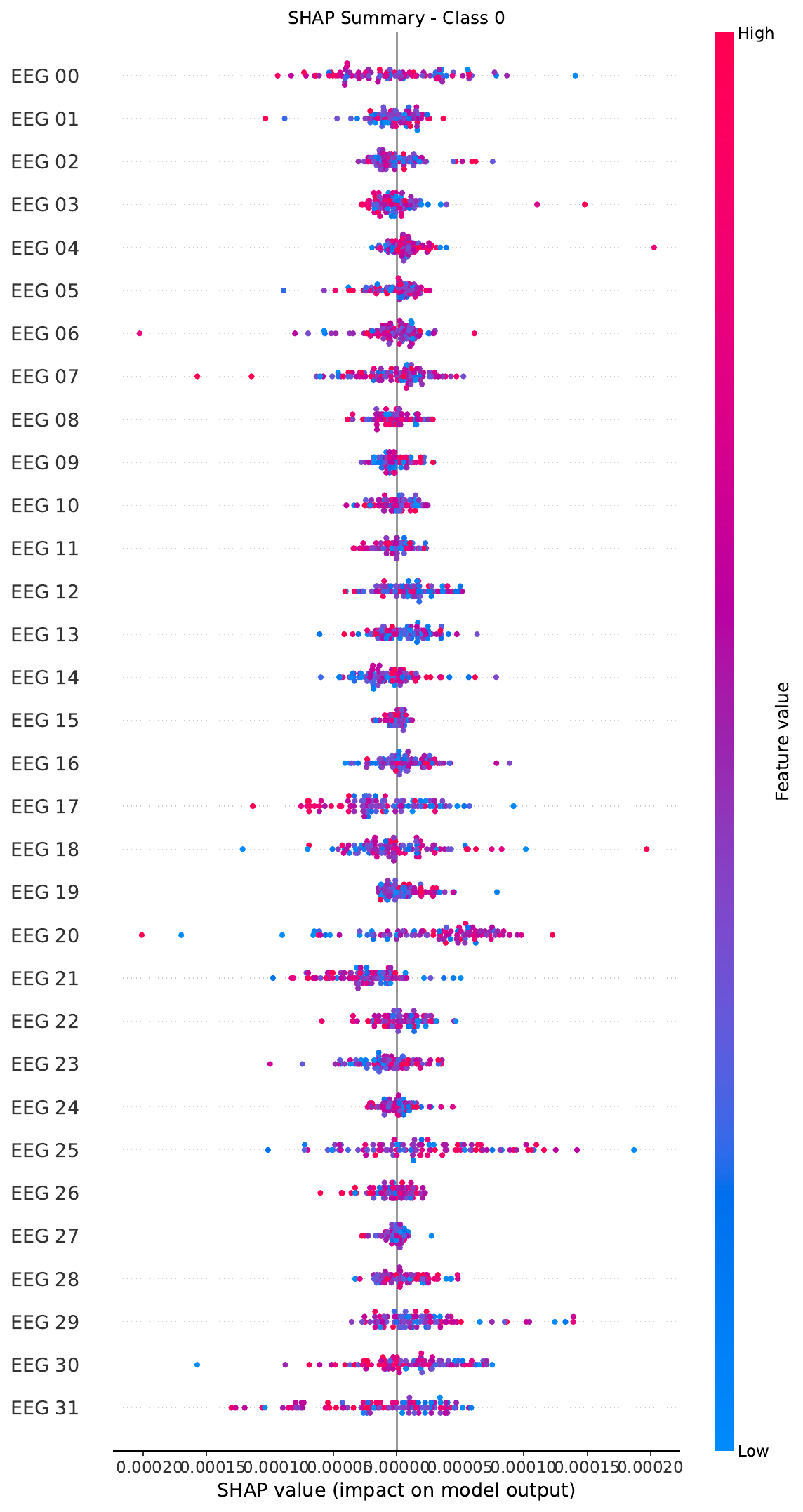}
    \caption{} 
    \label{fig_shap_0} 
  \end{subfigure}
  \begin{subfigure}[b]{0.5\linewidth}
    \centering
    \includegraphics[width=\linewidth]{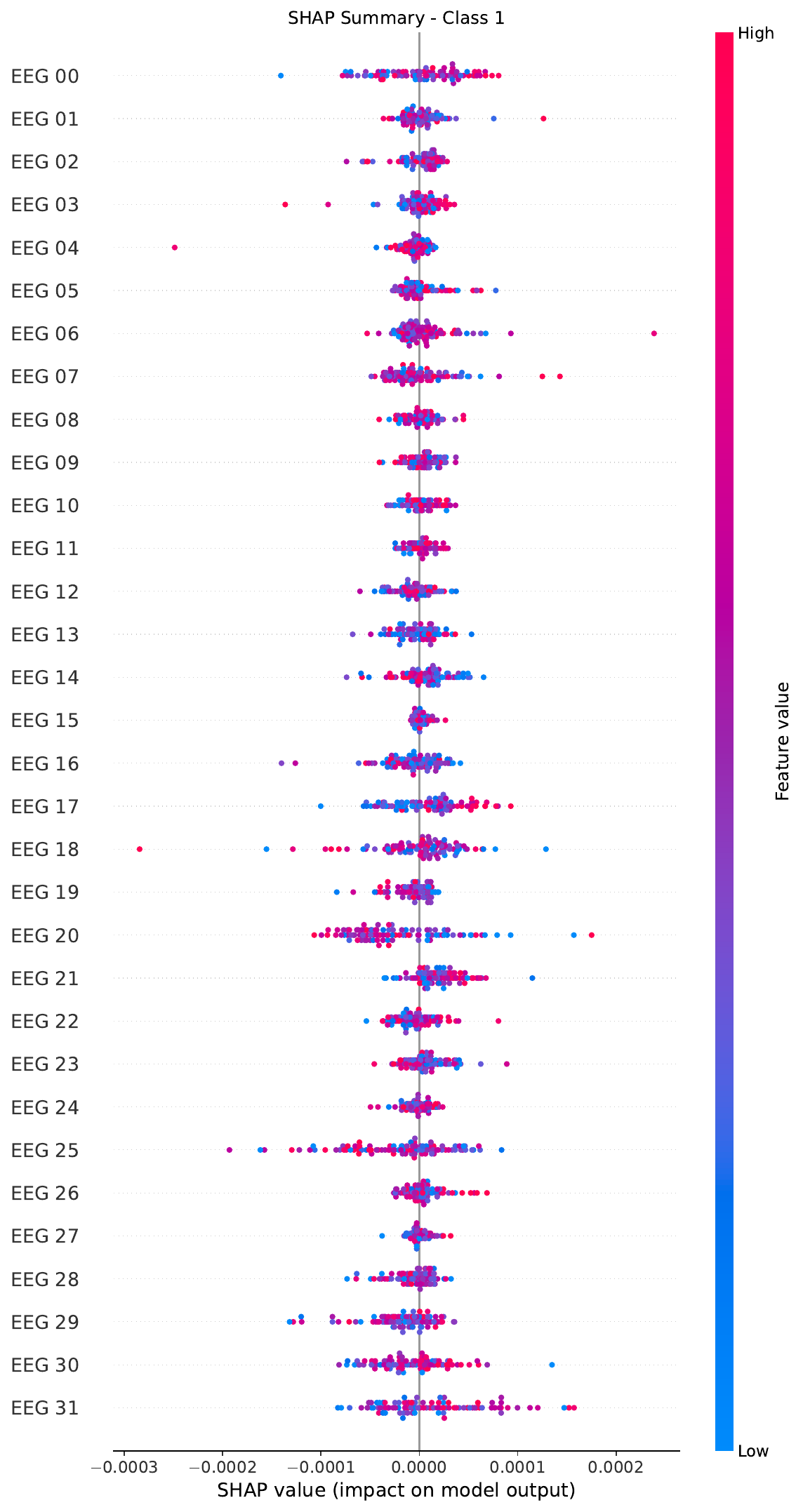}
    \caption{} 
    \label{fig_shap_1} 
  \end{subfigure}
  \caption{SHAP summary of EEG channel influence on prediction for (a) real and (b) synthetic samples. Top channels are more influential; colors show feature impact.}
  \label{fig_shap} 
\end{figure}

To gain a deeper understanding of the model's decision-making process, an interpretability analysis was conducted using SHapley Additive exPlanations (SHAP)~\cite{lundberg2017unified}, providing a unified measure of feature importance. As illustrated in \cref{fig_shap}, SHAP summary plots were generated to visualize the impact of individual EEG channels (EEG 00 to EEG 31) on the model's predictions for both real and synthetic samples. Each plot presents features ordered by their overall importance, from most influential at the top to least influential at the bottom. Each point on the plot represents the SHAP value for a specific instance in the dataset. The color of each point indicates the original feature value: red signifies a high feature value, while blue denotes a low feature value. \cref{fig_shap_0} displays the SHAP summary plot for real samples, revealing the channels that are most influential in classifying real EEG samples while \cref{fig_shap_1} shows the plot for synthetic samples. By comparing these two plots, any differences or similarities in feature importance between real and synthetic data can be identified, which is crucial for evaluating the model's robustness and generalization capabilities. This interpretability analysis is vital for confirming that the model is leveraging relevant physiological information from the EEG signals and not relying on spurious correlations.

\section{Conclusion}
\label{sec_conclusion}

The scarcity of high-quality, accessible raw EEG data was addressed in this study through the proposal of a conditional denoising diffusion model. Synthetic EEG samples were generated by augmenting noise to real data, creating similar but unique samples. It was demonstrated that classifier training is enhanced when synthetic data is used in conjunction with real data. For emotion classification, it was found that EEGNet and TSception outperformed SVM, with up to a 5.6\% accuracy increase observed when synthetic data was incorporated. It is confirmed that classification performance is improved by augmenting training with meaningful synthetic data, whereas it is reduced by random noise.

The proposed conditional diffusion model demonstrates promising potential for generating synthetic EEG data, offering a future direction for refined techniques in emotion recognition. However, computational cost and hyperparameter sensitivity are acknowledged limitations. Future work will build upon this foundation, with these challenges to be addressed to further refine diffusion-based EEG analysis. This includes the exploration of four-quadrant emotion recognition schemes for a more nuanced understanding of emotional states, and the evaluation of the model's performance with a wider variety of classifiers to confirm its generalizability across diverse machine learning paradigms. Furthermore, future studies will also explore model performance on DREAMER~\cite{katsigiannis2017dreamer} and SEED~\cite{zheng2015investigating} datasets to enhance the generalizability of the findings.




\section*{Bibliography}
\bibliography{references}








\end{document}